\documentclass[twocolumn,prb,aps,floatfix]{revtex4}

\usepackage{graphicx}
\usepackage{dcolumn}
\usepackage{bm}

\begin{document}

\preprint{APS/123-QED}

\title{On the possibility that STS ``gap-maps'' of cuprate single crystals
are dominated by k space anisotropy and not by nano-scale
inhomogeneity.}

\author{ J.R. Cooper}
\affiliation{Department of Physics, University of Cambridge, J. J.
Thomson Avenue, Cambridge CB3 OHE,U.K.} \altaffiliation{
Corresponding author; email address: jrc19@cam.ac.uk}

\date{\today}

\begin{abstract}
The results of a computer analysis of a simple 2D quantum
mechanical tunnelling model are reported. These suggest that the
spatial dependence of the superconducting energy gap observed by
Scanning Tunnelling Spectroscopy (STS) studies of single crystals
of the high $T_c$ superconductor Bi$_2$Sr$_2$CaCu$_2$O$_{8+x}$ is
not necessarily caused by nanoscale inhomogeneity. Instead the
spatial dependence of the STS data could arise from the momentum
($\textbf{k}$) dependence of the energy gap, which is a defining
feature of a $d$-wave superconductor. It is possible that this
viewpoint could be exploited to obtain  $\textbf{k}$ dependent
information  from STS studies.

\end{abstract}

\maketitle

\section{Introduction}

In recent scanning tunnelling spectroscopy (STS) experiments,
differential conductance ($G \equiv dI/dV$) versus voltage ($V$)
curves are measured at many thousands of points on the surface of
single crystals of the highly anisotropic cuprate superconductor
Bi$_2$Sr$_2$CaCu$_2$O$_{8+x}$ or Bi:2212 \cite{Lang, McElroy}. The
form of $G(V)$ at every point is roughly that expected from a
tunnel junction between a normal metal and a $d$-wave
superconductor. However, the superconducting gap parameter,
$\Delta$, determined from the peaks in $G(V)$, varies strongly and
continuously with position, from $20$ to $70$ meV with a typical
length scale of $1-2$ nm. It is thought that $G(V)$ is a direct
measure of the local quasi-particle density of states (DOS) and
therefore these ``gap maps'' are widely accepted as being
associated with nanoscale inhomogeneity, implying that the
superconducting gap and all other superconducting properties are
spatially inhomogeneous. It is difficult to reconcile this picture
with the results of other experiments such as NMR \cite{Alloul}
and heat capacity\cite{Loram}, where such gross inhomogeneity is
not detected.

 The energy ($E$) dependent quasi-particle DOS, $N(E)$ in a $d$-wave
superconductor with a $\textbf{k}$-dependent order parameter
$\Delta(\textbf{k})$ can be thought of as the sum of $s$-wave like
contributions from different directions in $\textbf{k}$-space,
i.e.,  $N(E)= \frac{2}{(2 \pi)^3} \int
\frac{dS}{\hbar|v(\textbf{k})|} \frac{E}{[E^2-
\Delta(\textbf{k})^2]^{1/2}}$, where $dS$ is a small element of a
constant energy surface in the normal state with electron velocity
$v(\textbf{k})$. The measured quantity $G(V)$ is proportional to
$\frac{2}{(2 \pi)^3} \int dE\int
\frac{dS}{\hbar|v(\textbf{k})|}P(\textbf{k})
\frac{E\delta(E-V)}{[E^2- \Delta(\textbf{k})^2]^{1/2}}$,  where
$P(\textbf{k})$ is the probability of tunnelling into  a state
$\textbf{k}$, so $G(V)$ only reflects the behavior of $N(E)$
precisely when $P(\textbf{k})$ is constant.
  For a defect-free \textit{planar} metal-insulator-superconductor tunnel junction,
   transverse momentum ($\textbf{k}_T$)  will be
conserved because of translational invariance. Also, because of
the exponential decay of the evanescent wave in the barrier, the
tunnelling current will be dominated by states with smaller values
of $|\textbf{k}_T|$, that lie within a ``tunnelling cone'' whose
angular width depends on the strength and width of the barrier
\cite{Wolf}. So, for \emph{planar} tunnelling into a
superconductor with a reasonably large Fermi surface,
$P(\textbf{k})$ is not constant, because final states with larger
values of $|\textbf{k}_T|$ will not be accessible. But for planar
tunnelling perpendicular to the conducting layers of a quasi-2D
material such as Bi:2212, $P(\textbf{k})$ is constant if the
quasi-2D Fermi surface is electron-like and exactly
circular\cite{Wei}, or if the ``tunnelling cone'' is wide enough,
i.e. $|\textbf{k}_T|$ is large enough, so that $P$ does not vary
much over the Fermi surface \cite{Wei}. At first sight, spatial
confinement of the initial electron states in the sharp STS tip
would be expected to give a larger spread in $\textbf{k}_T$, so
there would be an even greater tendency for $P$ to be constant for
$c$-axis tunnelling into Bi:2212 using STS. This seems to be
 the prevailing view in the field. While the model calculations reported
 here do show that for realistic STS tip sizes and tip materials, $\textbf{k}_T$ is
not conserved,  i.e. there is no well-defined refracted wave in
the quasi-2D metal, they also show that $P$ is not constant.
Instead it falls drastically as the in-plane wave number of
Bi:2212, represented by the parameter $|k_y|$ in the
 model, is increased.

  Recent STS studies  give clear
experimental evidence that  bigger gaps are observed in areas of
the crystal where there are more off-stoichiometric oxygen atoms
in the BiO-SrO layer \cite{McElroy}.  In a naive semi-classical
picture these  O$^{2-}$ ions would scatter the tunnelling
electrons, via the Coulomb interaction, giving them extra
transverse momentum. So
 there could be a tendency for a particular  in-plane direction of $\textbf{k}$
to have extra weight in $G(V)$ at a certain separation between the
STS tip and an O$^{2-}$ ion and hence a particular value of
$\Delta(\textbf{k})$ could be favored there. This provides a
mechanism for variations in the peaks in $G(V)$ as the STS tip is
moved across the surface of the crystal without invoking nanoscale
inhomogeneity.

The  model calculations reported here are consistent with this
semi-classical picture in that the strong $k_y$ dependence of $P$
mentioned above becomes much less marked when there are O$^{2-}$
ions near the STS tip. Although they do not give a completely
adequate description of the observed $G(V)$ curves, they seem to
capture the important physics and future microscopic calculations
of the $G(V)$ curves should take the present results into account.

\begin{figure}
\includegraphics[width=7.5cm,keepaspectratio=true]{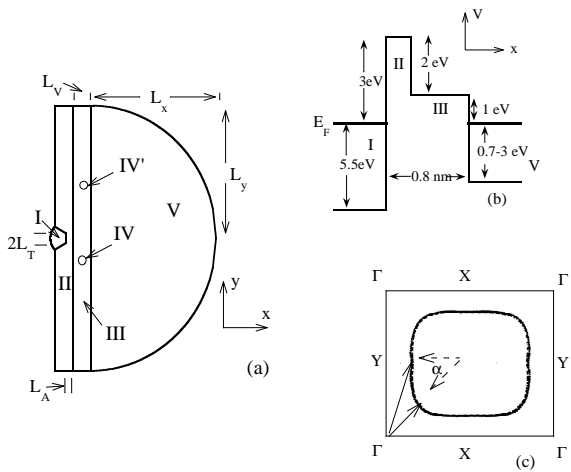}
\caption{(a) Spatial regions used for 2D tunnelling calculations,
\textrm{I} STS tip, \textrm{II} vacuum, \textrm{III} BiO-SrO
layer, \textrm{IV} and \textrm{IV'} oxygen ions, \textrm{V}
 Bi:2212. (b) Corresponding potential energies versus distance
($x$) at $y$ = 0. (c) Typical Fermi surface
cross-section\cite{Pickett}, with solid arrows marking the minimum
and maximum $\textbf{k}$ vectors considered here and dashed arrows
showing the angle $\alpha$ with the $(-\pi,0)$ direction (see
text).}
 \label{tunnelFS}
\end{figure}

\section{The model}
The spatial dimensions and energy barriers of the  2D quantum
mechanical tunnelling model are defined in Figures 1(a) and 1(b)
respectively. Region \textrm{I} corresponds to the STS tip, which
is the projection of a truncated cone with sides at $30$ degrees
to the surface normal, i.e. to the $c$-axis of the cuprate
crystal. The Fermi energy of $5.5$ eV in the tip corresponds to a
free electron wave vector of $12$ nm$^{-1}$ which is close to the
Fermi wave vector (${k_F}^{Au}$) of typical tip materials such as
Au or a Pt alloy. The vacuum gap  in region \textrm{II} was
generally taken to have a width $L_A$ = $0.24$ nm and a barrier
height of $3$ eV as deduced from  STS work \cite{Kitazawa}. The
boundary condition on the small line of length $2L_T$ represents
an incoming electron wave of the form
$\exp[ik_{F}^{Au}(x\cos\theta + y\sin\theta)]$ travelling at an
angle $\theta$ to the normal, i.e. on this line (where $x$=0) the
real and imaginary parts of the wave function, $\psi_R$ and
$\psi_I$ were set equal to $cos (k_{F}^{Au}y\sin\theta)$ and
$sin(k_{F}^{Au}y\sin\theta)$ respectively. The STS tip width
$2L_T$ was often fixed at $0.3$ nm but calculations were made for
a range of values between $0.15$ and $5$ nm. For $2L_T$  =  $5$ nm
(i.e. $10$ times the wavelength of the electrons in the Au tip)
the result expected for a planar junction, namely conservation of
transverse momentum was obtained. However for $2L_T \leq 0.6$ nm
there was no evidence for $\textbf{k}_T$ conservation, i.e. no
sign of a wave propagating at the appropriate angle in region
\textrm{V}.  The  angle $\theta$  was varied between $0$ and $30$
degrees, but since the number of initial states in the tip
increases as $\sin\theta$, Ref. \onlinecite{Wolf}, it was decided
that $25$ degrees was  appropriate for comparison with experiment.
This choice does not affect our conclusions.

\begin{figure}
\includegraphics[width=8.0cm,keepaspectratio=true]{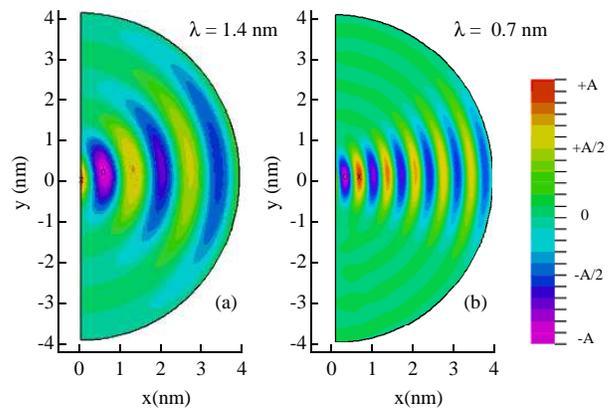}
\caption{(Color on-line). Contour plots in the $(x,y)$ plane of
real part of wave function $\psi_R$ in region \textrm{V} for
wavelengths of (a) 1.4 nm and (b) 0.7 nm, using the parameters
given in Fig. 1(b) and without any O$^{2-}$ ions. Here $x$ is
measured from the start of region \textrm{V}. The linear color
scale is also shown.} \label{contourplots}
\end{figure}

Region \textrm{III} corresponds to the BiO-SrO layer which is
taken to be $0.56$ nm wide \cite{Structure}. In most of the
calculations a barrier height of $1$ eV,  consistent with band
theory \cite{Pickett}, was used . This layer is now believed to be
insulating in view  of (a) the extremely large and non-metallic
$c$-axis resistivity of Bi:2212 crystals \cite{Watanabe}and (b)
various experiments on intrinsic $c$-axis tunnelling in mesa
structures, for example Ref. \onlinecite{Krasnov}. Region
\textrm{V}, where the potential energy $V(x,y)$ is constant,
corresponds to the bulk of the cuprate superconductor.
Representing the electronic structure of the cuprates by a free
electron model, which is effectively done here, is a drastic
approximation. However including an extra barrier to represent the
$c$-axis lattice potential, i.e. the next BiO-SrO layer, does not
alter the main conclusions because a defect-free planar barrier
would reflect semi-circular waves such as those shown  in Figure 2
without altering $k_y$. This point was checked  explicitly by
introducing an appropriate extra planar barrier in region
\textrm{V}. This had no effect on the   behavior shown later for
$P(k_y)$ in the absence of O$^{2-}$ ions,  but because of standing
waves there was more scatter in the results.

\begin{figure*}
\includegraphics[width=13.0cm,keepaspectratio=true]{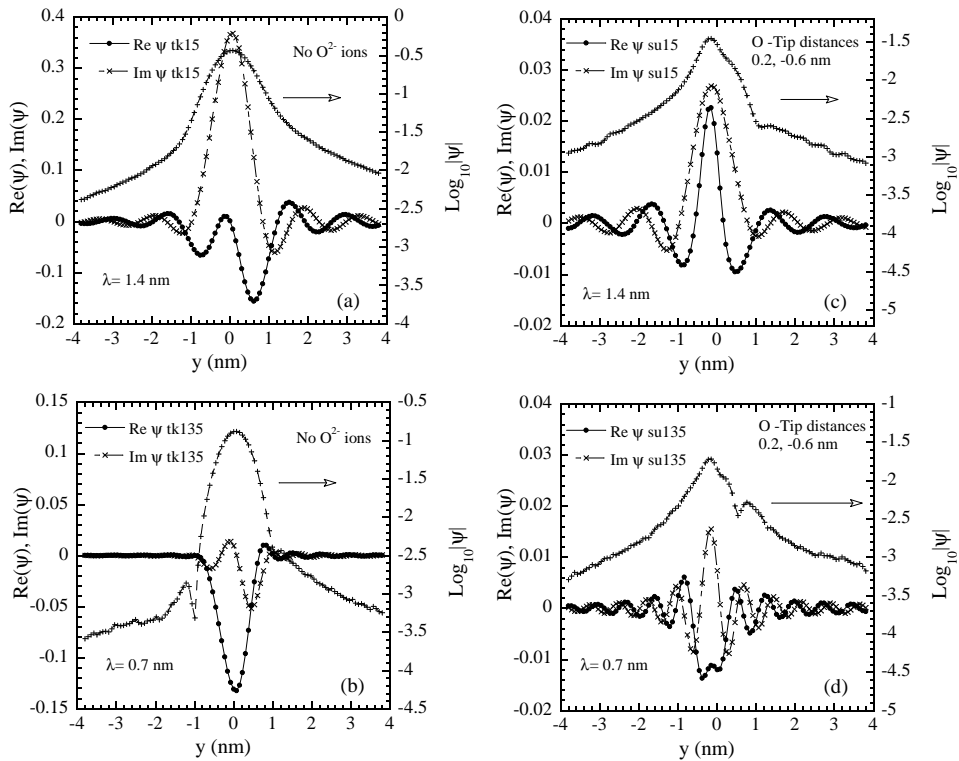}
\caption{Plots of the electron wave function ($\psi_R$, $ \psi_I$
and $|\psi|$) in a narrow strip of width $x$ =  $0.05$ nm
extending from $y$ = -3.8 to 3.8 nm, just inside region
 \textrm{V}, i.e. the first CuO$_2$ layer. Computed data for the
two limiting wavelengths are shown in (a) and (b) in the absence
of O$^{2-}$ ions and in (c) and (d) with O$^{2-}$ ions  at +0.2
and -0.6 nm from the center of the STS tip. Parameter values are
those in Figure 1 (b), with 2$L_T$ = 0.3 nm, $L_A$=0.24 nm and
$\theta$ (see text) = 25 degrees. }
 \label{psiplots}
\end{figure*}

The present model is therefore a useful starting point for
discussing the effect of oxygen ions in the barrier layer and for
discussing whether the STS technique does indeed give a true
$\textbf{k}$-space average of the electronic DOS. The value of the
Fermi energy $E_F$ in region \textrm{V} was varied over the range
$0.76$ to $3.04$ eV because this corresponds to free electron wave
numbers from $4.47$ to $8.94 $ nm$^{-1}$ and wavelengths between
$1.4$ and $0.7$ nm. These wave numbers are approximately equal to
the minimum and maximum in-plane values of $k_F$ for electron
states at the Fermi energy in the cuprates  shown in Figure 1(c).
In tunnelling problems one always considers the ``physical
electron'' as emphasized by Anderson in connection with
inter-layer tunnelling in the cuprates.

The small circular regions \textrm{IV} and \textrm{IV'} in Figure
1(a) correspond to spherically symmetric O$^{2-}$ ions with a full
outer shell situated $0.1$ nm from the CuO$_2$ layer.  They give
rise to a Coulomb field outside a typical atomic radius of $0.06$
nm. In the metallic region \textrm{V}, this field will be strongly
screened  by the mobile electrons (on a length scale of order
$k_{F}^{-1}$). As often done in tunnelling problems
\cite{Schmidlin} this screening is represented by an equal and
opposite image charge in region \textrm{V} giving a dipole field
from each non-stoichiometric O$^{2-}$ ion in the BiO-SrO layer.
For simplicity, dielectric screening in the BiO-SrO layer is
ignored since positive polarization charges at the \textrm{II-III}
boundary can then be neglected. However in order to compare with
experiments, where the mean spacing between O$^{2-}$ ions is
typically $0.8$ nm,  the Coulomb field was multiplied by a
Gaussian attenuation factor, $\exp(-(y-y_1)^2/\sigma^2)$ where
$y_1$ is the $y$ co-ordinate of the O$^{2-}$ ion and $\sigma$ =
$0.4$ nm. This was done because the tunnelling electrons will be
scattered by \emph{local} increases in the Coulomb potential above
a uniform background potential given by more distant O$^{2-}$
ions. The Coulomb potential inside region \textrm{IV} was set
equal to a constant, namely its value at the \textrm{III-IV}
boundary. For simplicity it was also assumed that the dipole
electric field does not extend into the vacuum region \textrm{II}
or into the metallic region \textrm{V}.

\section{Method and Results} A commercially available partial
differential equation (PDE) PC package\cite{Flex} was used to
solve the 2D Schr\"{o}dinger equations for $\psi_R$ and $\psi_I$
with the above potentials.  An example program attributed to
Backstrom\cite{Flex} and initial tests for the textbook problem of
tunnelling through a 1D potential barrier\cite{Schiff} were
helpful in setting the boundary conditions (BCs), details of which
are given in footnote \onlinecite{BCs}. The BCs on the
semicircular bounding surfaces shown in Figure 1(a) were chosen so
that outwardly propagating waves are favored and the absence of
significant standing waves was verified by comparing plots of
$|\psi|$ and $\psi_R$ along various lines in region \textrm{V}.
Most of the calculations were done for the geometry shown in
Figure 1(a) with $L_x$ = $L_y$ = $4$ nm but checks were made for
larger values of $L_x$, for a rectangular bounding region and also
for a cylindrically symmetric 3D case. The graphical output given
by the PDE program was typically in the form of contour plots of
$\psi_R$ such as those in Figures 2(a) and (b) or plots of
$\psi_R$ and $\log|\psi|$ along certain lines (not shown). Arrays
of $101$ x $ 101$ $\psi_R$  and $\psi_I$ values over various
user-defined regions of Figure 1(a) were also generated and these
were processed further using other commercially available PC
packages.

Figures 2(a) and (b) show examples of the contour plots for the
two extreme $k_F$ values mentioned above in the absence of any
impurities in the barrier. Note that the shorter wavelength waves
are more concentrated in the forward direction. As justified in
detail later, this leads to the key result of the present work,
namely $P(k_y)$ is not constant for tunnelling between a sharp STS
tip and Bi:2212 and therefore  the $G(V)$ curves will give a
distorted image of the DOS. The barriers are relatively weak so
that the width of the ``source'' along $y$ just inside region
\textrm{V} is considerably wider than that of the STS tip ($0.3$
nm). For the parameters given in Figure 1(b) and used in Figure 2,
the regions of high intensity below the STS tip and $0.2$ nm
inside the conducting region \textrm{V} have FWHM in $|\psi|$, of
$\delta y$ = $1.16$ nm and 0.84 nm, for $\lambda$ = $1.4$ and
$0.7$ nm respectively. The ratios $\lambda/\delta y$ differ by a
factor $1.5$ and this, together with the fact that $\lambda\simeq
\delta y$ and $\delta x$, must be be responsible for difference in
spreading out of the waves shown Figures  2(a) and 2(b). Another
important point is that, as mentioned in the previous section,
there is little asymmetry between positive and negative $y$
values, despite the fact that  the wave coming in to the STS tip
is at $25$ degrees to the surface normal, and therefore has
significant transverse momentum $k_y = k_F^{Au}\sin(25)=5.1$
nm$^{-1}$ along $y$. In contrast to the planar tunnelling case, no
refracted wave that conserves $k_y$ is produced because the STS
tip width, $2L_T$ = $0.3$ nm is too small, as shown later in
Figure 4(c).

Computed data are shown in more detail in Figures 3(a) to 3(d) as
plots of $\psi_R$, $\psi_I$ and $\log|\psi|$ versus $y$ at a mean
distance along $x$ of $0.25$ nm into region \textrm{V}. The data
shown are in fact the sum of $100$ $\psi$ values between $ x$ =
$0$ and $0.5$ nm, now measured from the beginning of region
$\textrm{V}$, for fixed values of $y$. A strip of this width
corresponds to the first CuO$_2$ bi-layer in
Bi:2212,\cite{Structure} but the same results are obtained for
different strips e.g. between $x$ = $0$ and $0.3$ nm. The $|\psi|$
plots are shown on a semi-logarithmic scale in order to see the
behavior at larger values of $y$.  There are no significant
oscillations in $|\psi|$ showing that the BCs do indeed produce
propagating and not standing waves. The periods of the
oscillations in $\psi_R$ and $\psi_I$ along $y$ are very close to
the two wavelengths used, $0.7$ and $1.4$ nm. Figures 3(a) and
3(b) show data  with no O$^{2-}$ ions in the BiO-SrO layer.
Comparison of the two figures shows that the shorter wavelength
has lower amplitude oscillations, by at least a factor $10$, over
most of the range of $y$ confirming the differences shown in
Figures 2(a) and 2(b). Figures 3(c) and 3(d) show the effect of
having one O$^{2-}$ ion at $y$ = $+0.2$ nm and another at $y$ =
$-0.6$ nm, a spacing typically observed in the experiments
\cite{McElroy},  the center of the STS tip being at $y$ = $0$. The
key point is that in the presence of
 O$^{2-}$ ions the amplitude is much less dependent on the wavelength.
This suggests  that STS does not give a true DOS average for
quasi-2D superconductors and that the presence of  O$^{2-}$ ions
allows states with larger values of $k_y$ to be accessed.

\section{Comparison with experiment}
 In order make an initial comparison with experiment the following
procedure was used.  The $\psi(y)$ data such as those shown in
Figures 3(a) to 3(d) were multiplied by $\exp(-\imath 2\pi
y/\lambda)$ or $\exp(\imath 2\pi y/\lambda)$ and integrated over
$y$ to give the (complex) numbers $a(k_y)$ and  $a(-k_y)$. These
are the quantum mechanical overlap integrals between an initial
electron state in the STS tip and  plane waves of wavelength
$\lambda$ propagating in the $\pm y$ directions in region
\textrm{V}. Hence the quantity $|a(k_y)|^2 + |a(-k_y)|^2$ is the
required probability $P(k_y)$ of electron tunnelling from the tip
into a state in region \textrm{V} with $|k_y|$ = $2\pi/\lambda$.
$P(k_y)$  was calculated for a range of $\lambda$ values between
$1.4$ and $0.7$ nm.  Since the Schr\"{o}dinger equation is linear
in $\psi$ and the solutions are obtained by matching $\psi$ and
$\nabla \psi$ at the boundaries it is plausible, but not proved,
that the same probability function $P(k_y)$ would also apply to
the physically realistic case of the Bi:2212 bands where, as shown
in Figure 1(c), there is a range of $|\textbf{k}|$ values with the
same (Fermi) energy. Typical behavior of $P(k_y)$ is shown in
Figures 4(a) and 4(b) for various positions of the STS tip
relative to one or two O$^{2-}$  ions respectively. Generally $P$
varies as $A \exp(-B|k_y|)$, where A and B are constants, but
especially when the spacing of the two ions is $\sim\lambda$,
there is some extra curvature in plots of $\log P$ vs. $k_y$ which
probably arises from two-beam interference effects. As shown by
the curve $\textit{ta}$ in Figure 4(b), two nearby O$^{2-}$ ions
reduce $P$ by 2-3 orders of magnitude, this is understandable
because the height of the BiO-SrO barrier is increased locally
from $1$ eV to approximately $5$ eV.  Figure 4(c) shows the
behavior of $P(k_y)$ for different STS tip widths ($2L_T$) ranging
from $0.15$ to $3$ nm in the absence of O$^{2-}$ ions. The
exponential behavior mentioned above persists up to $2L_T$ = $0.6$
nm, after which there is a sharper fall associated with  the
tendency towards conservation of $k_y$ for a planar junction.

\begin{figure}
\includegraphics[width=8.0cm,keepaspectratio=true]{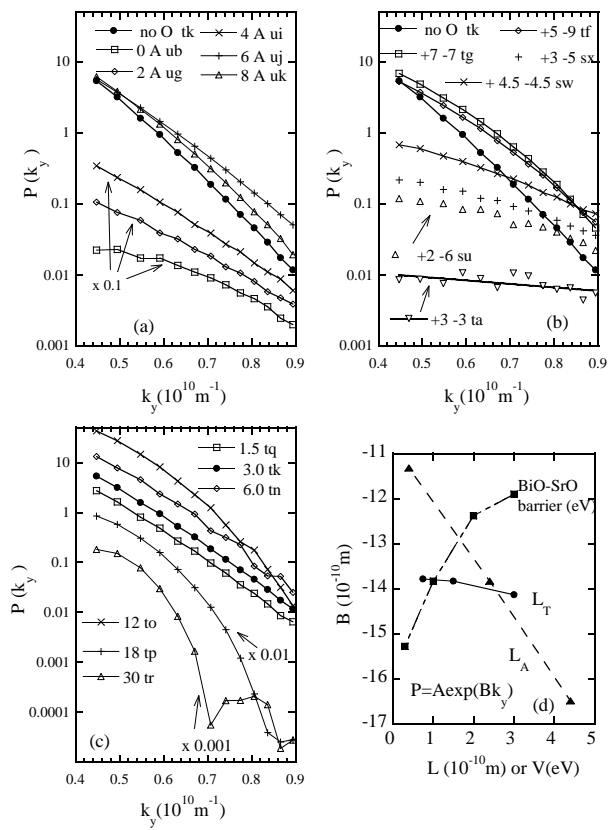}
\caption{Plots of $P(k_y)$, the probability weighting factor (see
text) versus $k_y$, (a) with no  O$^{2-}$ ions present (solid
black circles) and with one ion present at different distances, in
A$^o$ units (0.1nm)  from the STS tip. (b) with pairs of ions at
different distances, in A$^o$, and (c) with no ions present but
for various tip widths, 2L$_{\textrm{T}}$  in A$^o$. (d) Effect of
various parameters, the BiO-SrO barrier height in eV, the vacuum
gap L$_\textrm{A}$ in A$^o$, and the STS tip half-width
L$_\textrm{T}$ on the mean slope of log $_e$($P$) vs. $k_y$ in the
absence of O$^{2-}$ ions.} \label{probplots}
\end{figure}

 These $P(k_y)$ values were then used to calculate the corresponding $G(V)$ curves
expected in the superconducting state. Referring back to Figure
1(c), the contribution to $G(V)$ from a \textbf{k}-vector on the
Fermi surface was weighted by $P(|\textbf{k}|)$ where
$|\textbf{k}|$ is the distance from the $\Gamma$ point. Because of
the simplicity of the model, calculations were limited to the
region between the two solid arrows in Figure 1(c). The behavior
of $P(k_y)$  for different O$^{2-}$ environments shown in Figures
4(a) and (b)  means that small gap regions near the $d$-wave nodes
are heavily favored in the absence of O$^{2-}$  ions but that
scattering from O$^{2-}$ ions allows higher gap regions to be
accessed. Figure 4(d) shows that this conclusion is not strongly
dependent on  parameter values such as the STS tip width $2L_T$,
the BiO-SrO barrier height or the spacing between the STS tip and
the Bi:2212 surface ($L_A$). Some typical calculated $G(V)$ curves
are shown as un-normalized plots in Figure 5(a) and as normalized
ones in Figure 5(b). In generating these curves  the standard
weak-coupling $d$-wave form for the gap parameter,
$\Delta(\textbf{k})=\Delta\cos(2\alpha)$ was used, where $\alpha$
is the angle between $\textbf{k}$ and the $(-\pi,0)$ direction
shown by dashed arrows in Figure 1(c). Tunnelling data invariably
show some broadening of the $G(V)$ curves that is ascribed to a
finite quasi-particle lifetime ($\hbar/\Gamma$). In the Dynes
formula \cite{Dynes}
 used by Wei et al. \cite{Wei} $E$ is
replaced by $E- \imath\Gamma$ and $N(E)$ is given by the real part
of the usual expression. This formula should not be applied when
$\Gamma\sim\Delta$ \cite{Dynes}and therefore, in the present
calculations, moderate damping, namely $\Gamma = 0.1
\Delta(\textbf{k})$ was assumed for all $\textbf{k}$. In future it
might be possible to derive experimental values of
$\Gamma(E,\textbf{k})$ by appropriate fits to the STS $G(V)$
curves.
\begin{figure}
\includegraphics[width=8.0cm,keepaspectratio=true]{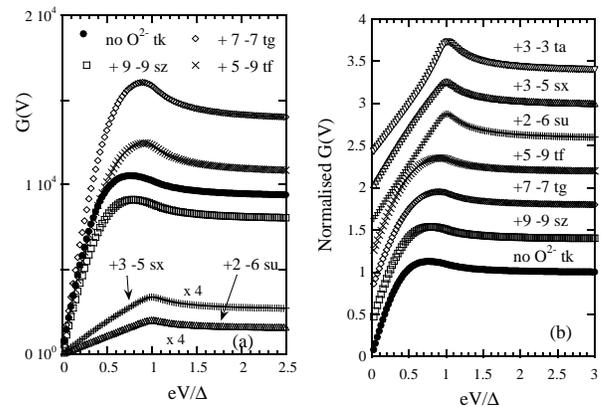}
\caption{(a) Calculated $G(V)$ curves for a $d$-wave DOS in the
same (arbitrary) units plotted versus $eV/\Delta$  for no O$^{2-}$
ions, curve $tk$, and pairs of O$^{2-}$ ions at various distances,
$y$ in A$^o$, from the STS tip at $y$ = 0. (b) Corresponding
$G(V)$ curves normalized to unity at $eV=3\Delta$ and displaced
vertically by multiples of 0.4.}

\label{G(V)plots}
\end{figure}
\section{Discussion}
 The un-normalized $G(V)$ curves shown in Figure 5(a) are
 similar to some of the published experimental data,
e.g. the first paper by Pan et al.\cite{Pan} where there is
clearly a strong reduction  in the magnitude of $G(V)$ in the
regions where the gap appears to be larger. It also noteworthy
that curves labelled \textit{tk}, \textit{tg}, \textit{tf} and
\textit{sz} seem to show very similar behavior at low voltages – a
feature that is also present in the experimental data, for example
Figure 3  of Ref. \onlinecite{Lang}.  The peaks in $G(V)$ vary
from $V$ = 0.7$\Delta /e$ to $\Delta /e$ . However even though
$P(k_y)$ varies drastically with $k_y$ when there is little
scattering by the O$^{2-}$ ions, the corresponding $G(V)$ curves
do appear to be more rounded than those observed experimentally,
especially recently \cite{McElroy,Lee}.

Within the present approach even higher $\textbf{k}$- space
selectivity may be needed to account for the experimental STS
data. There are several ways in which this might occur. Firstly in
the present model the band-structure of the BiO-SrO layer is
ignored. In a microscopic treatment one would consider tunnelling
to arise from virtual excitations into the BiO-SrO conduction
band.  This band is likely to have a narrow energy width and hence
for a given impact parameter the transverse momentum gained by
scattering from an O$^{2-}$ ion may well be better defined than in
the present model. Secondly the band structure of the CuO$_2$
layer is not included. Since this is a tight binding band
structure, Fermi surface states nearer the Brillouin zone
boundary, i.e. those with small values of the angle $\alpha$ in
Figure 1(c), will  contain significant components of states with
higher momenta, namely $\textbf{k} + \textbf{G}$ where
$\textbf{G}$ is a reciprocal lattice vector. This will give even
less weight to the states with higher values of  $|\textbf{k}|$.

 Thirdly the standard
``semiconductor model'' is used  to describe the tunnelling DOS of
a superconductor. The theory of Blonder, Klapwijk and Tinkham
(BTK)\cite{BTK}, which makes use of the Bogoliubov equations
rather than the Schr\"{o}dinger equation,  often provides good
fits to $G(V)$ curves of superconductors obtained with either
metallic point contacts or small-area tunnel junctions, despite
fact that it uses a 1D model. BTK theory contains some features
that could be relevant here, for example the quasi-particle
current transforms into a supercurrent over the coherence length
which in the present case is highly $\textbf{k}$-dependent. A
paper \cite{Kashiwaya} applying BTK theory to a $d$-wave
superconductor does conclude that $G(V) \sim N(V)$, but because a
1D model is used, $\textbf{k}_T$ is conserved, in contrast to the
present work.

Fourthly, the presence of a pseudogap would also reduce or
eliminate contributions to $G(V)$ from the regions of the Fermi
surface with larger values of $|\textbf{k}|$.  In a model used to
account for heat capacity data \cite{Loram2} there is a triangular
non-states-conserving pseudogap that ARPES experiments, e.g. Refs.
 \onlinecite{Ding} and  \onlinecite{Marshall}, suggest has greatest effect in the anti-nodal
directions. Introduction of a pseudogap would help in fitting the
V-shaped, apparently non-states-conserving $G(V)$ curves that are
often observed experimentally by STS \cite{McElroy,Lee,Cren1,Pan}.
Reports from other STS groups \cite{Cren1,Renner,Fischer}
suggesting that, especially for overdoped Bi:2212 crystals, the
normalized $G(V)$ curves can be uniform for linear scans over $10
- 20$ nm do not necessarily contradict the picture presented here,
since as shown in Figs. 4(a), 4(b) and 5(b) the normalized
$P(k_y)$ curves and hence the normalized $G(V)$ curves, can be
quite similar  for a variety of O$^{2-}$ spacings.

 Finally and perhaps most importantly, the
$\textbf{k}$-selectivity could be further increased by the
interference of scattered waves from different O$^{2-}$ ions.
There are some indications of this effect in the $P(k_y)$ curve
labelled \textit{tg} in Figure 4(b) where the STS tip is placed
symmetrically between two ions at a distance $\sim\lambda$ from
each. These effects could well be more marked in the realistic 3D
case where ``multiple beam'' interference is more likely, for
example  the scattered waves from $3$ or $4$ suitably spaced
O$^{2-}$ ions could interfere constructively for certain
directions of $\textbf{k}$ to give $9$ or $16$ times higher
intensity.   Since the positions of the O$^{2-}$ ions can be found
using STS\cite{McElroy} it might even be possible to locate such
structures and hence obtain better $\textbf{k}$-resolution in the
experiments. A faster PDE program would be needed to investigate
these 3D aspects using the present model. However the main result
reported here, namely the strong variation of $P$ with $k_y$ in
the absence of O$^{2-}$ ions and the weaker dependence in their
presence has been verified for one special 3D case.  This is the
cylindrically symmetric situation where the O$^{2-}$ ion is
positioned immediately below the STS tip and the electron wave
comes in at normal incidence.

\section{Summary and Conclusions}
A new way of interpreting the ``gap-maps'' observed in STS studies
of cuprate superconductors has been proposed that does not invoke
nanoscale inhomogeneity. In principle, perhaps when extended to
include real atomic orbitals and band structure, it can be used to
 obtain $\textbf{k}$-resolved information from STS data.

\textit{Note added on  26th Jan. 2007.}  Since submitting this
paper I became aware of analytical theory\cite{Tersoff} of the
scanning tunnelling microscope for a spherical STS tip. Equations
(4) and (9) of Ref. \onlinecite{Tersoff} lead to results that are
consistent with the present work for a single barrier without any
O$^{2-}$ ions. Neglecting terms with reciprocal lattice vectors,
$\overrightarrow{G}\neq 0$, gives a tunnelling probability
$P\propto \exp{[-2(\kappa^2 +
|\overrightarrow{k}_{||}|^2)^{1/2}|\overrightarrow{r}_0|]}$,
where\cite{Tersoff} $\kappa= \hbar^{-1}(2m\phi)^{1/2}$ is the
minimum inverse decay length, $\overrightarrow{k}_{||}$ the
in-plane electron wave-number, $\phi$ the work function (i.e. the
barrier height relative to $E_F$) and $\overrightarrow{r}_0$ the
distance between the center of curvature of the STS tip and the
sample surface. For a barrier height of 3 eV and
$|\overrightarrow{r}_0|$ = 0.8 nm, fitting the above formula to
$P$ = $A\exp{(Bk_y)}$ over the range of $k_y(\equiv
\overrightarrow{k}_{||})$ used here gives $B$ = -9.6 $10^{-10}$
m$^{-1}$. Bearing in mind that the present calculations were made
for a flat STS tip and a $2D$ model, this agrees well with the
data point in Figure 4(d), where $B$ = -12 $10^{-10}$ m$^{-1}$
when the BiO-SrO
 and vacuum barrier heights are both 3 eV, and the
width of the combined barrier is
 0.8 nm.

\section{Acknowledgements} This work forms part of a long-standing
collaboration with  J.W. Loram and J.L. Tallon who have provided
key insights and suggestions at all stages. Helpful suggestions
were received from C. Bergemann, who also supplied  the PDE
program, and T. Benseman.



\begin{thebibliography}{99}

\bibitem{Lang}K.H. Lang, V. Madhavan, J.E. Hoffman, E.W. Hudson, H.
Eisaki, S. Uchida and J.C. Davis, Nature (London), {\bf415}, 412
(2002).

\bibitem{McElroy} K. McElroy, Jinho Lee, J.A. Slezak, D.-H. Lee, H. Eisaki, S. Uchida and
J.C. Davis, Science, {\bf309}, 1048 (2005).

\bibitem{Alloul} J. Bobroff, H. Alloul, S. Ouazi, P. Mendels, A. Mahajan, N.
Blanchard, G. Collin, V. Guillen, and J.-F. Marucco, Phys. Rev.
Lett. {\bf89}, 157002 (2002).

\bibitem{Loram} J.W. Loram, J.L. Tallon and W.Y. Liang, Phys. Rev. B {\bf69}, 060502(R) (2004).




\bibitem{Wolf}E.L. Wolf, Chapter 2, \textit{Principles of Electron Tunneling
Spectroscopy} (Oxford University Press, Oxford, 1985).

\bibitem{Wei}J.Y.T. Wei, C.C. Tsuei, P.J.M. van Bentum, Q. Xiong,
C.W. Chu and M.K. Wu, Phys. Rev. B {\bf57}, 3650 (1998).

\bibitem{Pickett} W.E. Pickett, Rev. Mod. Phys. {\bf61}, 433
(1989).

\bibitem{Kitazawa} G. Kinoda, T. Hasegawa, S. Nakao, T. Hanaguri, K. Kitazawa, K. Shimizu,
J. Shimoyama and K. Kishio, Phys. Rev. B {\bf67}, 224509 (2003).

\bibitem{Structure} R.M. Hazen in \textit{Physical Properties of High
Temperature Superconductors II} (World Scientific, Singapore,
1990), D.M. Ginsberg (Ed.), pages 121-198.

\bibitem{Watanabe} T. Watanabe, T. Fujii and A. Matsuda, Phys.
Rev. Lett., {\bf79}, 2113 (1997).

\bibitem{Krasnov} V. M. Krasnov, A. Yurgens, D. Winkler, P. Delsing, and T. Claeson, Phys. Rev. Lett.
{\bf84}, 5860  (2000).

\bibitem{Flex} FlexPDE 3, PDE Solutions Inc. Antioch, CA.

\bibitem{Schiff}L.I. Schiff, Chapter V \textit{Quantum Mechanics}
(McGraw-Hill, New York,1955).

\bibitem{BCs}Using the syntax of ``FlexPDE 3", the BCs on all
non-vertical straight lines in Fig.1 were Natural($\psi_R$) and
Natural($\psi_I$) = 0, i.e. normal derivatives were set to zero
there. The BCs on all vertical lines except the STS tip were
NOBC($\psi_R$) and NOBC($\psi_I$), i.e no BCs were applied there,
nor were any applied to the small circles  representing the
O$^{2-}$ ions. The BC on the STS tip representing an incoming wave
at an angle $\theta$ to the normal is explained in the text. The
BCs on the large semi-circular bounding region were
Natural($\psi_R$)= $-\frac{2\pi}{\lambda}\psi_I$ and
Natural($\psi_I$)= $\frac{2\pi}{\lambda}\psi_R$ representing an
outgoing wave. The BCs on the small arc of the STS tip were the
same but with $\frac{2\pi}{\lambda}$ replaced by $k_F^{Au}$.

\bibitem{Schmidlin} F. W. Schmidlin, Journ. Appl. Phys. {\bf37}, 2823
(1966). Note that in this work the barrier  height is
\emph{decreased} by \emph{positively} charged ions.
\bibitem{Dynes} R.C. Dynes, V. Narayanamurti and J.P. Garno, Phys.
Rev. Lett. {\bf41}, 1509 (1978).
\bibitem{Pan}S.H. Pan, J.P. O'Neal, R.L. Badzey, C. Chamon, H. Ding, J.R. Engelbrecht, Z. Wang, H. Eisaki,
 S. Uchida, A.K. Gupta, K.-W. Ng, E.W. Hudson, K.M. Lang and J.C. Davis, Nature (London) {\bf413}, 282 (2001).

\bibitem{Lee} Jinho Lee, K. Fujita, K. McElroy, J.A. Slezak, M. Wang, Y. Aiura, H. Bando,
 M. Ishikado, T. Masui, J.-X. Zhu, A.V. Balatsky, H. Eisaki, S. Uchida and J.C. Davis,
  Nature (London) {\bf442/3}, 546 (2006).

\bibitem{BTK} G.E. Blonder, M. Tinkham and T.M. Klapwijk, Phys.
Rev. B {\bf25}, 4515 (1982).

\bibitem{Kashiwaya} S. Kashiwaya, Y. Tanaka, M. Koyanagi, and K.
Kajimura, Phys. Rev. B {\bf53}, 2667 (1995).

\bibitem{Cren1} T. Cren, D. Roditchev, W. Sacks, J. Klein, J.-B.
Moussy, C. Deville-Cavellin and M. Lagu\"{e}s, Phys. Rev. Lett.
 {\bf84}, 147 (2000).

\bibitem{Loram2} J.W. Loram, K.A. Mirza, J.R. Cooper and J.L.
Tallon, J. Phys. Chem. Solids {\bf59}, 2091 (1998).

\bibitem{Ding} H. Ding, T. Yokoya, J. C. Campuzano, T. Takahashi, M. Randeria, M.
R. Norman, T. Mochiku, K. Kadowaki, J. Giapintzakis,  Nature
{\bf382}, 51 (1996).

\bibitem{Marshall} D.S. Marshall, D.S. Dessau, A.G. Loeser, C.-H.
Park, A.Y. Matsuura, J.N. Eckstein, I. Bozovic, P. Fournier, A.
Kapitulnik, W.E. Spicer and Z.-X. Shen, Phys. Rev. Lett. {\bf76},
4841 (1996).

\bibitem{Renner} Ch. Renner, B. Revaz, J.-Y. Genoud, K. Kadowaki, and \O. Fischer, Phys. Rev. Lett.
 {\bf80}, 149 (1998).

\bibitem{Fischer} B.W. Hoogenboom, K. Kadowaki, B. Revaz and \O.
Fischer,
 Physica C {\bf376-380}, 15 (2003).
 \bibitem{Tersoff} J. Tersoff and D.R. Hamann, Phys. Rev. B {\bf31}, 805 (1985)
 \end{thebibliography}
\end{document}